# Competition and Globalization: Brazilian Telecommunications Policy at Crossroads*

**Clelia Piragibe**
**Advisor, Anatel Brazil**

## Introduction

The current pattern of competition of the Brazilian's telecommunication market was defined by a regulatory reform implemented in the second half of the nineties. Before that, there were only a state holding company – Telebras –, which controlled 23 state operators, and a long distance carrier – Embratel. (Section 1) each one with a monopoly in their respective markets.

The demands for more competition in the Brazilian telecommunications market started to make echoes around the early 90s, following the same "liberalization" and "deregulation" rhetoric observed worldwide. The telecommunications' regulatory reform, discussed in Section 2, promoted the privatization of Telebras System and fostered competition in the Brazilian market.

The privatized operators have achieved remarkable results under a duopoly system stricted supervised by Anatel – the new regulatory agency created by the 1997's General Law of Telecommunications (Section 3). The called LGT defined the regulatory framework of the new system.

Notwithstanding, the regulatory Brazilian scheme is now at a crossroads. From 2002, an open market approach will be implemented in telecom arena. In Section 4, we will discuss the possible changes in the pattern of competition in those markets and discuss changes, from the firms' strategy perspective, and also the main influences coming from the international scenario under the World Trade Organization' system.

**1.   Background**

Following the same pattern of most of the countries, until 1997 the State supplied and legally monopolized all telecommunications services in Brazil. That situation only contrasted with the United States' experience, where ATT had never been publicly owned or assured a legal monopoly.

In 1996, the Telebras System (a state holding company[1]) was the largest telecommunications company in Latin America. 28 operators were under its umbrella, with 88,000 employees.[2] Along the years, Telebras utilized a self-financing approach (the candidate user had to buy companies' shares in order to apply for a telephone line), combined with resources coming from National Fund of Telecommunications (FNT). That fund also supported the establishment of Embratel, in 1965, which came to explore long distance telecom services in Brazil.

The process of technology acquisition, adaptation and improvement by Telebras and Embratel evolved to a process of increasing local content and even technological developments along the eighties. Meanwhile, the System faced increasing bottlenecks to finance the huge investments necessary to their expansion. Macroeconomic reasons (massive government deficits), combined with microeconomic ones (low tariffs, the limitations of self-financing) explained the increasing limits to the System's growth.

---

[1] The Brazilian government controlled 5O.O4% of common shares and 21.45% of total capital at the end of 1996. 4O% of total preferred shares were held by international shareholders.

[2] Only four operators did not belong to the holding: CRT, Sercomtel, Ceterp. and CTBC.

When tariffs were concerned, the self-financing scheme had perverse effects on consumers. The high price paid to apply to the telephony service was a strong barrier to entry to poor people. Even when the self-financing was abolished, the candidate to a telephone line had to pay more than US$1.000 (converted in shares of Telebras' subsidiaries) and to wait until two years to have his terminal installed!

As far as local service was subsidized (in the form of low tariffs) by long distance calls and data communication, the elite[3] that accessed the services had a double privilege. From one size, they could protect their investments by negotiating the acquired shares in the stock market at higher prices. From the other size, they could get low prices by the use of the telecom services in Brazil.

The tariffs were also an element of dispute between local operators and Embratel, which operate long distance calls and data communications. As far as the last had the monopoly of the most profitable services, the embattles by shared revenues and services were constant.

The demands for more competition in the Brazilian telecommunications market, however, started only in the 90s. In 1997, the unsatisfied demand by telephones was estimated in 25 million terminals, in contrast with the 14.7 million access lines in service. The teledensity in Brazil in the same year (9.1) was below Argentina (21.9 and 17.9), Chile (15.4) and Mexico (9.7).

---

[3] In mid 90s, 98% of the Telebras' subscribers had a monthly income of more than US$ 1.000. At that moment, the minimum wage in Brazil was below US$100.

## 2. The Telecommunication Reform: 1995 to 1998

In the administration of Mr. Sergio Motta, head of the Brazilian Ministry of Communications from 1995 to 1998, the issues of Telebras System's privatization and the liberalization of the market were formally introduced. That position came in the focus of a wider liberal reform promoted by the government of Brazil under Mr. Cardoso' platform.

The main goals were services' universalization and competition between operators. The chosen formula was to privatize the telecom operators, keeping the government control on telecom services through regulation. Meanwhile, neither the financial resources nor the legislation were available at that moment.

Between 1995 and 1997, the major laws concerning Brazilian's privatization and liberalization of the telecommunications market came into effect:

1) The *Constitutional Amendment* (8/1995) authorized the break of Telebras' monopoly;

*2)* The *Minimum Telecommunications Law* (7/96) started the "flexibilization" of the model, opening to the private sector new concessions of mobile services, satellite services, data communications and value-added services

3) The *General Telecommunications Law* (7/97) was the masterpiece to the reform, defining the guidelines of an autonomous, original and open regulation to the sector. The fundamental principles established by the LGT were:

   1) The organization of telecommunications services according to two basic principles (universalization and competition);

2) The privatization of the federal companies (those ones under Telebras umbrella and Embratel).

3) The reorganization of telecommunications services market according to two basic principles (universalization and competition);

4) The creation of the regulatory agency (National Agency of Telecommunications, Anatel).

Regulatory authority has shifted in fundamental ways. Responsibility for the implementation of setorial policies, services and networks' regulamentation and inspection were taken from the Ministry of Communications to Anatel (Table 1). Issuance of authorizations for services in the public regime (subordinated to universalization and continuity principles, tariff control and so on) and authorizations for services in the private regime, monitoring of radiofrequency's spectrum and space, control of economic infractions were among the main roles of Anatel as well.

**Table 1**:

Telecommunications' Regulatory Reform in Brazil - 1997

|  | Before Reform | After Reform |
|---|---|---|
| **Policy' Formulation** | Ministry of Communications | Executive & Legislative |
| **Regulator** | Ministry of Communications/Telebras | Anatel |
| **Operator** | Telebras, Embratel | Private operators |
| **Owner** | Government, shareholders | Private shareholders |

**Source**: Anatel (2000)

A general restructuring plan was implemented before the privatization of the Telebras' System, as far as the idea of a mere change to a private monopoly was not acceptable:

1) Huge investments in modernization of the Telebras' System were implemented, between 1995 and 1997 (US$ 20 billion), through digitalization of the services, tariff's rebalance, and so on;

2) The state operators were spliced in two business units: one for fix telephony and other for mobiles;

3) The fixed telephony units passed to the umbrella of three regional fixed telephony holdings (responsible for local and intra-regional long distance services) and one long distance international and inter-regional company (see Figure 1);

4) The mobile units were combined in eight holding companies to provide services under the called "A band" (see Figure 2).

The Brazilian model was unique when comparable to other Latin America's experiences. The three-region approach targeted the integration of Brazil's North and Midwest according to the government's regional development policies. The creation of a competitive environment was driven to foment opportunities to attract investments and develop technology. Finally, the goal was to maximize the sale value of the telecommunications companies subject to the achievement of previous objectives. The profile of the Regional Fixed Line Holding Companies soon before their privatization is showed below.

**Figure 1:**
Brazil: Regional Markets for Fix Telephone Operators, 1998

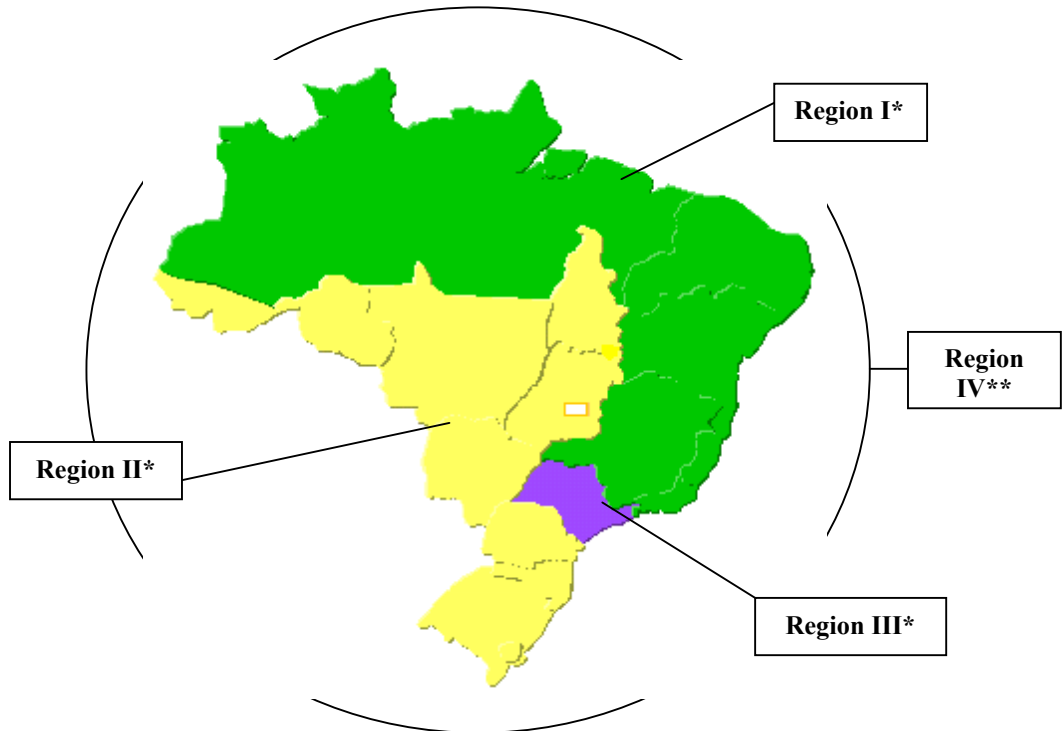

*Local and intra-regional
** Long distance

**Figure 2:**

Brazil: Mobile Market, "A Band" and "B Band" – 1998

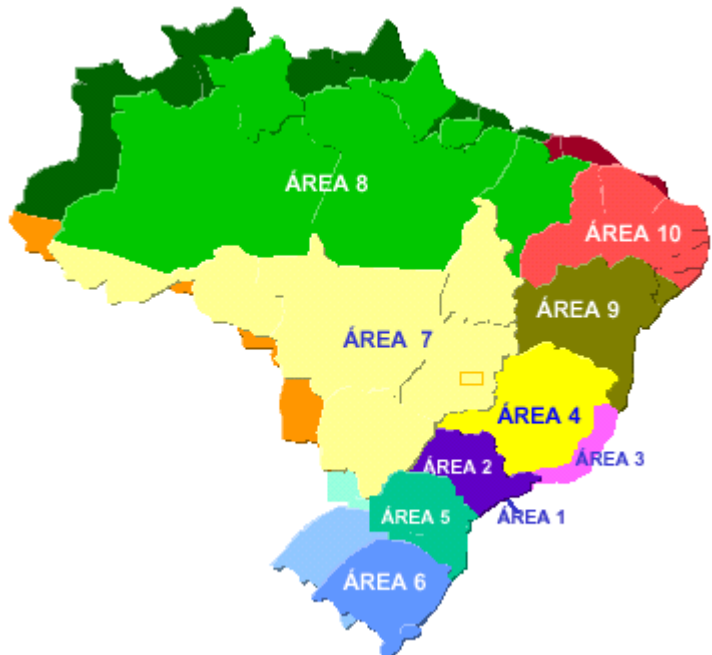

**Source**: Anatel

**Table 2:**

BRAZIL - Operational and Economic Profile of the New Telecom

Holding Companies before Privatization (1997)

|  | **TELESP** | **TELEST** | **TELESUL*** |
|---|---|---|---|
| **Lines in Service** | 5,704,239 | 5,642,452 | 2,742,713*** |
| -rank in LA | 3 | 2 | 6*** |
| **Employees**** | 24,162 | 37,106 | 19,724 |
| **Access Lines in Service Penetration** | 14,7% | 6.5% | 7.2% |
| % **of Brazil's GDP** | 36% | 39% | 25% |
| **1995 GDP per capita** | US$6,411 | US$2,740 | US$4,086 |
| % **of Brazil's popul.** | 22% | 54% | 24% |
| % **of Pop. Growth (91/96)** | 1.6% | 1.2% | 3.6% |

*Does not include CRT   **includes cellular   ***if CRT is included, access lines in service would total 3,694,939 and the company would rank 4th in Latin America.

**Source:** Restructuring and Privatization of the Telebras System: Overview of the Implementation Program (1997).

Meanwhile, as far as Sao Paulo – the richest state of Brazil – was defined as a single area, the model had a bias in terms of economic and demographic density. In fact, one of the main proposals of the privatization was to attract major international telecom companies to the Brazilian market, through the perspective of the huge growth of the local market and the absence of restrictions to the participation of the foreign capital in the acquisition of those companies.

In the beginning of 1997, the mobile market was open to private investments through the "B band'" bid. Ten regions were defined in order that the new operators could compete with the established ones ("A Band"). As a result, the Brazilian government earned more than US$ 8 billion with the auction (see Table 3), more than two times the defined minimum price.

Table 3

Brazil: Final Result of the Mobile Telephony' Auction, 1997 ("B Band")

- US$ Million -

| Area | Consortium | Minimum Price(A) | Paid Price (B) | B/A (%) |
|---|---|---|---|---|
| 1 | BCP | 600 | 2,646 | 341.25 |
| 2 | TESS | 600 | 1,326 | 121.16 |
| 3 | ATL | 500 | 1,508 | 201.28 |
| 4 | MAXITEL | 400 | 0,520 | 30.0 |
| 5 | Global Telecom | 330 | 737 | 134.52 |
| 6 | TELET | 330 | 334 | 1.36 |
| 7 | AMERICEL | 270 | 338 | 25.37 |
| 8 | NORTE BRASIL TEL. | - | 60 | - |
| 9 | MAXITEL | 230 | 250 | 8,69 |
| 10 | BSE | 230 | 555 | 141,55 |
| TOTAL | | 3,700 | 8,274 | 123,65 |

**Source:** Anatel' Site

In October 1997, the National Agency of Telecommunications – Anatel – was created. After that, the next step of telecom's reform in Brazil was the privatization of the Telebras' System. In order to achieve its mission, Anatel implemented some basic instruments to deal with the fixed telephone system:

1) *General Plan of Outorgs (PGO)*

1.1 According to the PGO, only the fixed telephone system would be explored under a public regime (or even under a public/private one); all others services were under a private regime;

1.2 An initial duopoly structure would be established em four areas (three licenses for local services and one national long distance services);

1.3 In each area, a "mirror company" would be created to compete with the incumbent ones to be privatized;

1.4 All new licenses would be auctioned: weighted score combining price with coverage and expansion targets would select winners.

2) *General Plan of Universalization Goals (PGMU)*

Only the incumbent firms must achieve the goals of the PGMU. Under the Plan, the target number of fixed terminals, at the end of 2001, should be 33 million, two times the 1997's figure. The Plan also target to increase the availability of local services according to the size of the cities and the teledensity of public phones per state and at country level (around 1 million at the end of 2001).

3) *General Plan of Quality Goals (PGMQ)*

Accordingly, this Plan defined the performance indicators related to better technical and operational standards to be achieved by the incumbent operators along the time.

Those instruments were essential steps prior to the privatization and also to the long-term environment of free competition planned to be implemented after 2001. The concession contracts established by the Agency with the incumbent operators – who define rights and obligations of the telecom companies – includes all parameters and indicators from the PGMU and PGMQ.

Along the transition period prior to 2001, several restrictions were defined as well. Each operator – the incumbent firm and its respective "mirror" company would be restricted to their regional and service market until they achieved their universalization and service quality's targets. In order to foment increasing competition, as well, each controller could only own shares of one single operator, in each services' category (for example one fixed line operator and one mobile operator).

In order to compete in new telecom markets in Brazil, from 2001 on, the operators must anticipate the 2003's targets (including PGMU and PGMQ ones) established in their concession contracts. Wisely, the regulatory agency defined the "carrot" which has motivated the operators to fulfill their universal services and quality obligations until the end of 2001.

**Table 4:**
Brazil: Privatization Auction of Telebras System, 1998
(US$ Million)

| Company | Consortium | Minimum Price (A) | Paid Price (B) | B/A (%) |
|---|---|---|---|---|
| Telesp Participacoes (fix) | Telefonica de Espanha, Portugal Telecom, Iberdrola, Banco Bilbao y Viscaya | 3,520 | 5,783 | 64.29 |
| Tele Norte Leste Participacoes (fix) | Andrade Gutierrez, Inepar, Funcef, A.D.Leite e Alianca da Bahia | 3,400 | 3,434 | 1.00 |
| Tele Centro Sul Participacoes (fix) | Telecom Italia, Opportunity | 1,950 | 2,070 | 6.15 |
| Embratel Participacoes (fix) | MCI | 1,800 | 2,650 | 47.22 |
| Telesp Celular Participacoes | Portugal Telecom | 1,100 | 3,558 | 226.18 |
| Tele Sudeste Celular Participacoes | Telefonica de Espana, Iberdrola, Itochu and NTT | 570 | 1,360 | 138.6 |
| Telemig Celular Participacoes | Telesystem, Pension Funds and Opportunity | 230 | 756 | 228.70 |
| Tele Celular Sul Participacoes | Globo, Bradesco and Telecom Italia | 230 | 700 | 204.00 |
| Tele Nordeste Celular Participacoes | Pension Funds, Opportunity and Telesystem | 225 | 660 | 193.30 |
| Tele Centro-Oeste Celular Participacoes | Grupo Beldi (Splice) | 230 | 440 | 91.30 |
| Tele Leste Celular Participacoes | Telefonica de Espana, Iberdrola | 125 | 428,8 | 242.20 |
| Tele Norte Celular Participacoes | Pension Funds, Opportunity and Telesystem | 90 | 188 | 108.90 |
| TOTAL | | 13.470 | 22,042 | 63.74 |

**Source**: Anatel' site

Once the Government, with the support of several consultancy firms, developed the desired structure of the new telecom market in Brazil, step by step, the basic conditions to the privatization process were achieved. The auction of the former state companies occurred in July 1998, and their main results are showed in Table 4.

In the first half of 1999, the definition of the "mirror" companies that would start to compete with the privatized incumbents was completed. Four new private operators entered in the Brazilian telecom market (see Table 5), and later some local auctions have been realized by the regulator to define smaller operators into specific "niche" markets (those companies were named "little mirrors").

**Table 5:**

Brazil: Main Characteristics of Telecom 'Mirror' Companies

-Fix Market-

| Region | Name of the Company /Year Start operation | Controller | Technology/market |
|---|---|---|---|
| I | Vesper S.A./2000 | Velocom (bought % of Bell Canada ); Qualcomm | WLL/ local, intra-regional |
| II | Global Village (GVT)/2001 | | Local, intra-regional |
| III | Vesper S.P./2000 | Velocom (bought % of Bell Canada); Qualcomm | WLL/ local, intra-regional |
| IV | Intelig/2000 | National Grid, France Telecom, Sprint | Local, intra-regional |

**Source**: several sources

As we can see, the 'mirror' companies are quite young and, together, they only control less than 7% of the number of fixed lines in the Brazilian telecom market.[4] The local market is still almost monopolized by incumbent firms. The market where stronger competition was reached until the moment is the intra-regional long distance calls, once until four companies are competing there (local operator, Embratel and their mirrors).

---

[4] In mid-2001 the three mirrors represent around 3 million terminals from a total of 43.3 million fix access lines intalled in Brazil.

## 3. Recent Developments in the Brazilian Telecommunications Market

The 'digital economy' represents today 10% of the Brazilian GDP. Together, the 200 largest private companies in the field of information/communications in Brazil had total sales of US$ 55 billion in the year 2000.

The largest revenues came from telecom companies (US$ 30.6 Billion). Among the twelve firms that got sales beyond one billion dollars in that year, seven came from telecommunications market. Telemar – the only incumbent that is controlled by Brazilian shareholders – was the leader with US$5.5 billion sales (see Table 6). The number of direct jobs by the telecom service operators reached more than 300.000 people in 2000, and the direct investments by those companies in the Brazilian market achieve US$ 7 billion dollars.[5]

**Table 6**

Brazil: The Top Ten Telecommunications Firms in 2000

| Company | Sales A (US$M.) | Profits B (US$M.) | Margin % B/A | Employees | Controller' Origin |
|---|---|---|---|---|---|
| Telemar (f) | 5,551 | 363 | 6.54 | 39,297 | Brazilian |
| Telefonica (f) | 5,025 | 752 | 14.97 | 56,151 | Spanish |
| Embratel (f) | 4,581 | 295 | 6.44 | 12,000 | American |
| Brasil Telecom (f) | 3,093 | 210 | 6.78 | 23,642 | Italian |
| Telesp Cel. (m) | 1,884 | 78 | 4.13 | 4,336 | Portuguese |
| Motorola (e) | 1,504 | N/A | N/A | 3,100 | American |
| Tele SE Cel. (m) | 1,014 | N/A | N/A | 1,326 | Spanish |
| Nokia (m) | 941 | N/A | N/A | 1,260 | Finland |
| BCP (m) | 853 | -182 | -21.34 | 1,908 | Brazilian |
| Tele CO Cel (m) | 578 | 66 | 11.45 | 2,397 | Brazilian |

f= fix operator     m= mobile operator     e= equipment maker
**Source:** Info Exame 2001

---

[5] The investments in communications in Brazil represented 24% of the total foreign investment until June 2001 (estimated in US$ 9,345 Billion ), followed by financial transactions (16.21%) and chemical industry (9.13%)

The four original incumbents (coming from the fix telephony market) represent 60% of the total sales of the sector (or US$ 18 billion). Three mobile incumbents are also among the largest firms (with sales of US$ 3,5 billion). The only new comer in the list (operating in the "B Band") shows huge losses in that period. If we combine the losses of the three mobile operators, which operate in "B Band" in Rio-Sao Paulo ax (the largest markets in Brazil), they will achieve US$ 500 million in 2000.

The growth of the Brazilian telecommunications market has been quite fast after the privatization. As far as Brazil is the largest Latin-American country, in terms of population and size of the economy, the stock of fixed and mobile followed those trends in recent years.[6] In 1999, Brazil showed a leader position in terms of the advances of fixed terminals worldwide (with a 25% growth compared with the previous year).

The number of fix access lines installed in Brazil increased more than three times between 1994 and 2001 (see Figure 3), reaching 43.3 million in mid-2001.[7] Telefonica – the incumbent that operates in Sao Paulo (the largest market in Brazil) - is responsible for 12.5 million lines only in that state.

The increase of the mobile market was already faster (see Figure 4). The number of cellular terminals reached 26 million in mid-2001. In the year 2000, the mobile's growth rate was above 50%, following the market *boom* between 1995/99, when the number of mobiles almost double each year in Brazil.

An important characteristic of the Brazilian telecom growth in recent years was the deeper penetration of the telecom services among poor people. Until 1994, more than 80% of the fix telephones belonged to the rich and middle class in Brazil.

---

[6] Brazil represents around 38% of the Region's GDP, and controls 37% of the fix access lines and 40% of mobiles. In contrast, Argentina shows a smaller proportion in both telecom indicators, and Mexico follows the trend with the mobile market. Data quoted by Wohlers, Marcio (2001).
[7] In 1994, Brazil had only 13.3 million fix telephone terminals already installed.

**Figure 3**

Brazil : Evolution of the Fix Telephone Lines in Operation – 1995/2001

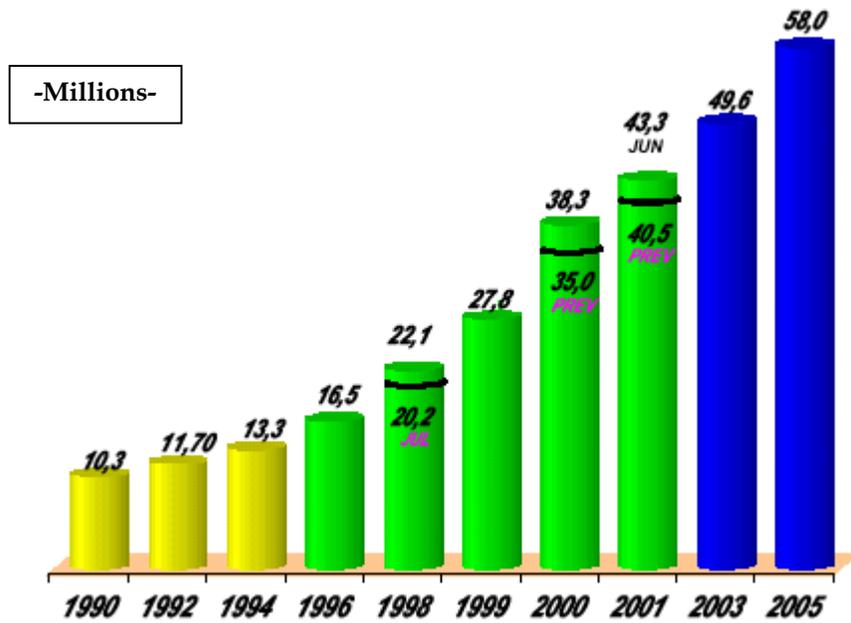

**Figure 4:**

Brazil: Evolution of Cellular Mobile Services, 1995/2001

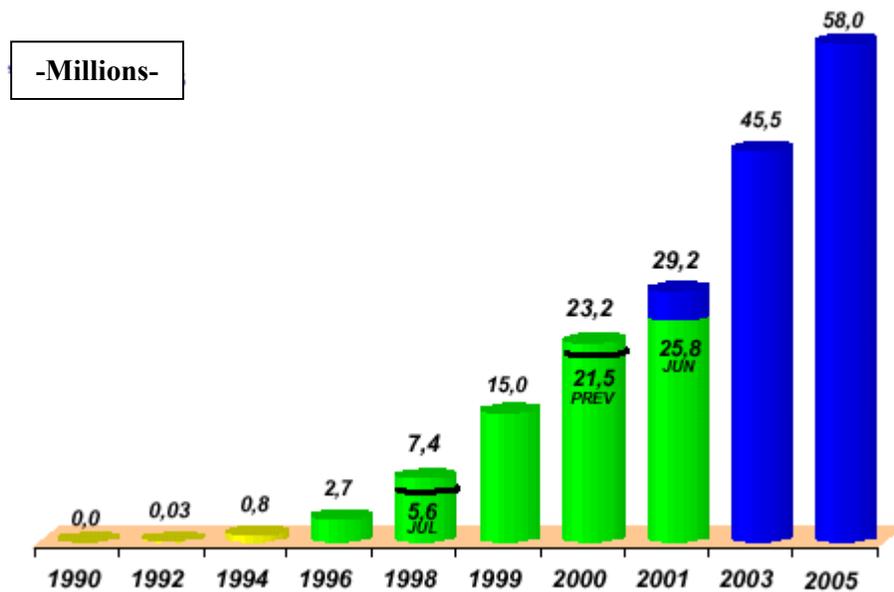

**Note: 2001 (2nd sem), 2003 and 2005, forecast**

**Source:** Anatel

Today, one quarter of the poorer families had a telephone set (see Figure 5). The public telephones along those years growed three times, reaching more than 1.000.000 terminals in mid-2001. The wide spread of mobiles among the poor in Brazil is quite explained by the huge growth of pre-paid cellular services (that represents 59% of the handsets in operation today in the country).

It is important to remember that corporative services (classified among private services), and particularly data communications services, are offered in an environment of strong competition in Brazil. Anatel gave more than 200 licenses to those operators, but most of the licensed companies are still providing services for their own use. Two to three dozen firms are concentrated in large cities and even in rich neighborhoods providing specialized services.

Meanwhile, some companies are reaching a strong growth in that area. AT&T Latin America` revenues increased 1,218%, from US$2.5 million, in 1999, to US$ 33 million in 2000. The company operates not only in Brazil, but also in Argentina, Peru, Chile and Colombia. It offers broadband services (including last mile) to corporative clients.[8]

The large incumbents – like Telemar, Embratel and Telefonica are also doing huge investments in this profitable market (like IP network), and promoting the capilarization of those services in their respective areas of concession. Even mirror companies like Vesper are diversifying its operations towards corporative users.

---

[8] The firm has more than 1,000 corporative users among very large corporations operating in Brazil – like GM, AOL, Acer and Souza Cruz. Another target is its participation in the Brazilian Payment System (SPB), which will establish a real time link among all banks, including Bovespa stock market.

**Figure 5:**
Brazil : Homes with Telephone According to Personal Income

**Region I:**

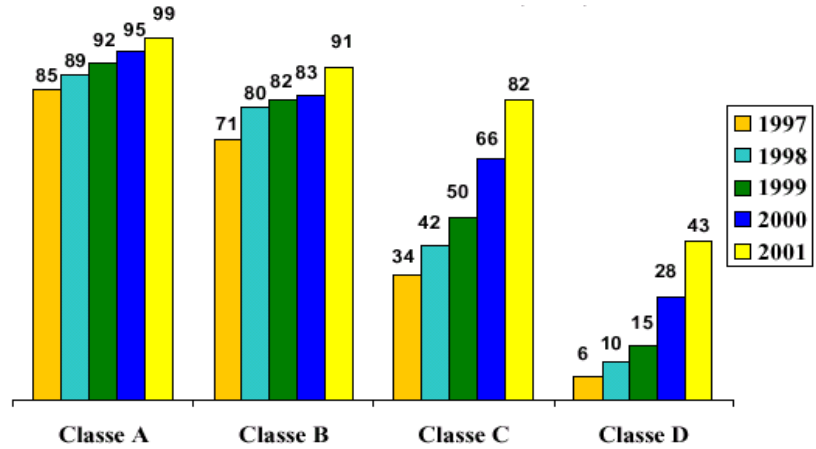

**Region II:**

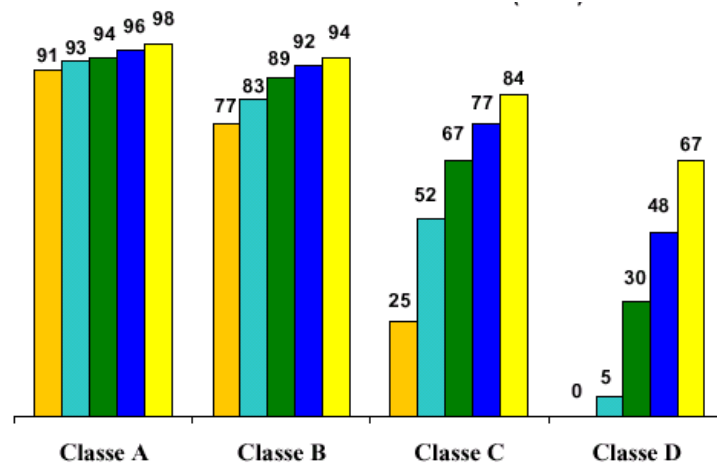

**Region III:**

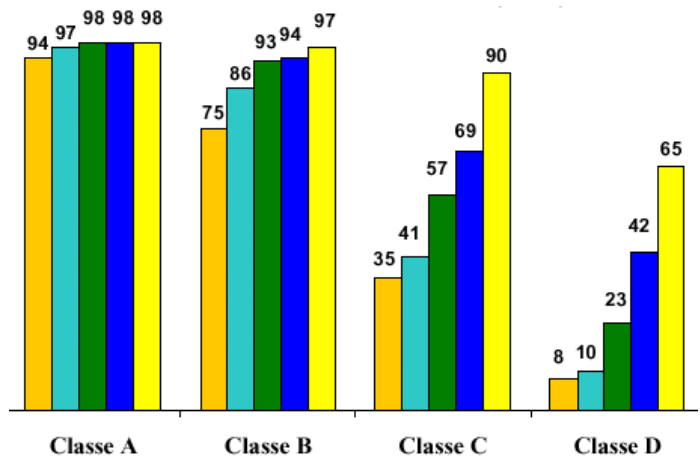

**Note:** 2001, forecast
**Source:** Anatel